# Extending Firewall Session Table to Accelerate NAT, QoS Classification and Routing

Mahmoud MOSTAFA, Anas ABOU EL KALAM, Christian FRABOUL

*Université de Toulouse, INPT-ENSEEIHT, IRIT-CNRS, 2 rue Camichel Toulouse – France*
*{firstname.lastname}@enseeiht.fr*

*Abstract* — security and QoS are the two most precious objectives for network systems to be attained. Unfortunately, they are in conflict, while QoS tries to minimize processing delay, strong security protection requires more processing time and cause packet delay. This article is a step towards resolving this conflict by extending the firewall session table to accelerate NAT, QoS classification, and routing processing time while providing the same level of security protection.

## 1. INTRODUCTION

Many firewall security mechanisms have evolved to mitigate the ever continuously increasing number of network attacks. Securing open and complex systems have become more and more complicated. If in addition we should take QoS requirements into account, the problem becomes more complicated and necessitates in-depth reflexions.

Router is a key component in the internet. Its main function is to control data packet flow and determine an optimal path to reach the destination. However, as the networking technology has evolved, much new functionality have been added and implemented in the router. In our context, besides routing function; we are also interested in other functionalities such as NAT, QoS, and stateful packet filtering.

Firewall is the primary defense perimeter to protect networks. Firewall technology had evolved from stateless packet filtering toward stateful packet filtering implemented in network routers. In order to be able to trace connection state; SFP builds a session table (also called state table). This session table makes SFP faster and more secure. Some implementations have extended the session table to include NAT mapping information.

In this work we find that adding QoS and routing information to the session table is a natural extension that will speedup performance and enhance router scalability and availability.

The remainder of this paper is organized as follows: in Section 2 we provide an overview of the security edge router's main functions. Section 3 presents session table architecture and processing; Then, Section 4 presents our integrated session table architecture and processing. Finally, we draw some conclusions and perspectives for future work in Section 5.

## 2. ROUTER MAIN FUNCTIONS

*Routing*

Routing is the process of selecting paths in a network along which to send network traffic [1]. Routing directs packet forwarding, the transit of logically addressed packets from their source towards their ultimate destination through intermediate nodes. The routing process usually directs forwarding on the basis of routing tables which maintain a record of the routes to various network destinations. Thus, constructing routing tables, which are held in the routers' memory, is very important for efficient routing. Routing table may be configured manually or dynamically utilizing routing algorithm such as OSPF [2]. Fast routing table lookup is an important requirement to implement high performance router. A lot of researches have been conducted to speedup routing table lookup [3], [4].

*Network Address Translation (NAT)*

**NAT** is an IETF [5] standard that enables a local area network (LAN) to modify network IP addresses and ports numbers in headers of datagram packets (in transit across a traffic routing device) for the purpose of remapping a given address space into another. One of the main objectives of NAT is to solve the scalability problem when the number of IP addresses allowed to access the external network is limited. From the security point of view, NAT more or less hides internal private network addresses from outsiders, enforces control over outbound connections, and restricts incoming traffics [6]. The NAT table is the heart of the whole NAT operation, which takes place within the router as packets arrive and leave its interfaces. Each connection from the internal (private) network to the external (public-Internet) network, and vice versa, is tracked and a special NAT table is created to help the router determine what to do with all incoming packets on all of its interfaces. Again, NAT table lookup plays an important role in enhancing router performance.

*Quality of Services (QoS) Processing*

QoS is a heavily loaded term with many different meanings depending upon the specific context. IETF [7] has defined QoS as nature of the packet delivery service provided, as described by parameters such as achieved bandwidth, packet delay, and packet loss rates.

The main goal of QoS is to provide priority treatment. A QoS policy should identify what priority level will be given to each traffic flow. After that, classification algorithms [8] can be used to inspect each packet and mark it with its associated priority level. High priority traffic such as VoIP should be served before non-priority one such as e-mail or FTP packets. To achieve this goal packets are placed in queues (waiting for processing) according to its priority levels. Queues represent locations where packets may be held (or dropped). Packet scheduling refers to the decision process used to choose which packets should be serviced or dropped. Buffer management refers to any particular discipline used to regulate the occupancy of a particular queue. Packets will be placed in different queues according to their priority levels. Afterwards, schedulers will pick packet to be served according to their priorities. The most important objectives of scheduling are computational efficiency and fairness [9].

*Packet Filtering*

Firewalls are network devices that filter network traffic at one or more of the seven ISO network model most commonly at the network, transport, and application levels [10].

Packet Filtering Firewalls (PFs) were the first generation of firewalls. They are basically screening routers [11] that control the flow of data in and out of a network by looking at certain fields in the packet header: *Source IP Address, Destination IP Address, Protocol identifier, Source port number, and Destination port number.*

The **PF** inspects all incoming and outgoing packets and applies the specified policy (e.g., drop or accept the packets).

**PF** was considered as an efficient, fast, and cost effective solution since a single router can protect an entire network. However, **PF** has a lot of limitations: it is based on IP addresses without any authentication, it depends on port number for identifying communicating applications and this is not a reliable indicator because many current protocols such as network file system (NFS) uses varying port numbers. It cannot defend against man in the middle attacks and forged packets with spoofed IP addresses. But the most important limitation is the difficulty of writing correct filters [12] for complex and permanently evolving systems. Generally, filtering rules are far from providing perfect security against holes in the **PF**.

*Stateful Inspection Packet Filtering Firewall (SPF)*

While **PF** works by statically inspecting each packet against the rule set, **SPF** works not only by inspecting the packet headers but also by correlating the incoming traffic to the earlier outgoing requests [13]. Basically, **SPF** builds dynamic session/state table to record relevant information of each communication to trace the validity of each packet in these connections. **SPF** dynamically opens and closes ports according to the connection needs, in this way it makes network management easier. The following section describes the structure of the session table.

3. SESSION TABLE ARCHITECTURE AND PROCESSING

Fig. 1. Shows the general architecture of the session table; the first five fields are <src-addr, src-port, dst-addr, dst-port, IP-p> used to identify a unique session , it is the same fields used by stateless firewall and QoS to classify traffic; it is called SID (session ID or selector). State field is used to store the state of this session and time field contains session timeout [14].

Fig. 2. Represents casual stateful packet filtering processing in router based implementation. For incoming packet, first, NAT translation will be performed to make the necessary mapping between external and internal addresses; then the session table will be searched, if an entry is found in the session table the packet will be inspected to ensure its conformance to the session state. If there is no entry in the session table, this means that the packet is the first

one in this session so, the packet will be validated against the filtering rules. If it is allowed to pass, an entry in the state table will be added. After that, QoS classification will be performed and the DSCP (differentiated service code point) priority value will be set in the ToS "Type of Service" (also called Traffic Class in IP v6) field in the packet IP header to tell core routers how to treat this packet [15]. Finally routing table will be looked up to determine the next hop and then the packet is transmitted.

| Src-aadr | Src-port | Dst–addr | Dst–port | IP-proto | State | Time |
|---|---|---|---|---|---|---|

Fig. 1. General architecture of session table.

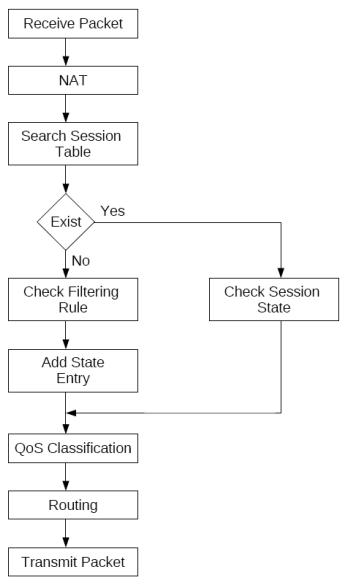

Fig. 2. Casual stateful packet filtering processing.

Not that all of the previously mentioned operations need tables' lookup and most of them search multiple fields in the table to find the appropriate entry. Reducing this lookup time is an important goal to enhance performance [16]. For this reasons some firewall implementation such as NetBSD PF [17] merged NAT and session state information in the session table.

## 4. OUR INTEGRATED SESSION TABLE ARCHITECTURE AND PROCESSING

Our goal is to increase security edge router processing capacity and enhance its scalability and availability. In order to achieve this goal, we merge all the needed information to perform SPF, NAT, QoS classification and routing in an integrated session table. This will make all the needed information available in only one search process in short session table, which is a great saving in processing time.

Fig. 3. Shows the architecture of our integrated session table; due to column wide space limitation the table is divided into two parts.

| NAT information | | | | | |
|---|---|---|---|---|---|
| Lan addr | Lan port | Gwy addr | Gwy port | Ext addr | Ext port |

| Stateful filtering info. | | | QoS | Routing info. | |
|---|---|---|---|---|---|
| IP-Proto | State | Time | SDCP | Ext-next-hop | Lan-next-hop |

Fig. 3. Our integrated session table architecture.

The first part of the table contains NAT information necessary to perform mapping between private and public addresses. Lan-addr and lan-port are the internally private addresses and port number while gwy-addr and gwy-port are the NATed publicly available address and port number. Finally, ext-address and ext-port are external communicating host IP address and port number.

For stateful filtering: < Lan-addr, lan-port, ext-addr, ext-port, IP-proto> constitute session ID and the session state and time out are stored in state and time fields.

For QoS classification the SID will be used to do classification and the QoS priority value will be stored in the DSCP field.

For routing table ext-next-hop will be used to send the packet to a destination outside the protected network. While, Lan-next-hop will be used to send the packet to a destination inside the protected network.

Fig. 4. Represents our integrated session table processing. For incoming packet, the session table will be searched, If there is no entry in the session table, this means that the packet is the first one in this session so, first, NAT translation will be performed to make the necessary mapping between external and internal addresses; then the packet will be validated against the filtering rules. If it is allowed to pass, an entry in the state table will be added. After that, QoS

classification will be performed and the DSCP will be added to its field in session table. Finally routing table will be searched to obtain the next hop values.

If an entry is found in the session table the packet will be inspected to ensure its conformance to the session state and all the needed session processing will be performed in one shot without further research overhead, as all the needed information is available from the single lookup in the short session table. This is clearly a great enhancement which save processing time and increase performance.

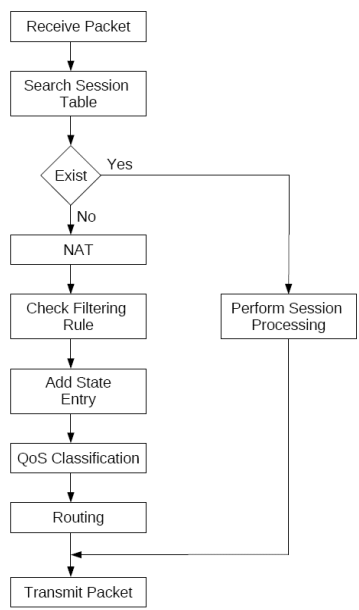

Fig. 4. Our integrated session table processing.

## 5. CONCLUSION

In this article, we presented the structure of our integrated session table that allow security edge router to performer the needed session processing (i.e. stateful filtering, NAT, QoS classification and routing). The use of our integrated session table produce great saving in router processing time and enhance its availability and scalability as it will be able to serve more traffic flows. Currently we are working in modifying NetBSD PF firewall kernel implementation to implement our session table architecture and processing. The implemented integrated session table will be tested and performance test results will be analyzed. The final implementation will be integrated in a QoS cable integrated security gateway which assures high level of security protection and high availability for time critical traffic.

REFERENCES

[1] Huitema, Christian. Routing in the Internet, Second Ed.. Prentice-Hall 2000.
[2] J. Moy, OSPF Version 2, IETF RFC 2328, April 1998.
[3] Jinxian Lin, Huimin Li, "Optimization of the Routing Table Lookup Algorithm for IPv6,", in: Second International Conference on Genetic and Evolutionary Computing, 2008. WGEC '08.
[4] T. Chiueh and P. Pradhan, "High performance IP routing table lookup using CPU caching," in Proc. of IEEE INFOCOM'99, New York, NY, 1999.
[5] Egevang, K. and P. Francis, "The IP Network Address Translator (NAT)", IETF RFC 1631, May 1994.
[6] Smith, M.; Hunt, R.; "Network security using NAT and NAPT" in 10th IEEE International Conference on Networks, 2002. ICON 2002.27-30 Aug. 2002 Page(s):355 - 360
[7] Shenker, S. and J. Wroclawski, Network Element Service Specification Template. RFC 2216, September 1997
[8] David E. Taylor, "Survey and taxonomy of packet classification techniques", ACM Computing Surveys. 2005.
[9] John Evans and Clarence Filsfils, "Deploying IP and MPLS QoS for Multiservice Networks: Theory and Practice"; Morgan Kaufmann, 2007
[10] Kenneth Ingham and Stephanie Forrest, "A History and Survey of Network Firewalls", Technical Report, TR-CS-2002-37, University New Mexico, 2002.
[11] Zwicky, E. D.; Cooper S. and Chapman D. B.: "Building Internet Firewalls", Orielly & Associates Inc., 2nd Edition, June 2000
[12] Al-Shaer, E.; Hamed, H.; Boutaba, R. and Hasan, M.: "Conflict Classification and Analysis of Distributed Firewall Policies", In IEEE Journal on Selected Areas in Communications, Volume 23, No. 10, pp. 2069 – 2084, October 2005.
[13] Siyan, Karanjit and Hare, Chris, "Internet Firewalls and Network Security", Indianapolis: New Riders Publishing, 1995
[14] Xin Li, Zhenzhou Ji, Mingzeng Hu; "Session Table Architecture for Defending SYN Flood Attack"; in 7th International Conference Information and Communications Security, ICICS 2005.
[15] Nichols, K., S. Blake, F. Baker, and D. Black, Definition of the Differentiated Services Field in the IPv4 and IPv6 Headers. RFC 2474, December 1998.
[16] D.E. Taylor, "Survey and taxonomy of packet classification techniques", ACM Computing Surveys 37(3), 2005, 238-275.
[17] NetBSD Packet Filter information, available at: < http://www.netbsd.org/docs/network/pf.html >